# *Quantum Finite One-Counter Automata*

Maksim Kravtsev[*], Department of Computer Science, University of Latvia, Raiòa bulv. 19, Riga, Latvia, e-mail maksims@batsoft.lv

(May 27, 1999)

**Abstract.**
In this paper the notion of quantum finite one-counter automata (QF1CA) is introduced. Introduction of the notion is similar to that of the 2-way quantum finite state automata in [1]. The well-formedness conditions for the automata are specified ensuring unitarity of evolution. A special kind of QF1CA, called simple, that satisfies the well-formedness conditions is introduced. That allows specify rules for constructing such automata more naturally and simpler than in general case. Possible models of language recognition by QF1CA are considered. The recognition of some languages by QF1CA is shown and compared with recognition by probabilistic counterparts.

## 1. Introduction

Quantum computation has proved to be of great interest for current and future researches. In the quantum computation theory many counterparts of the basic constructions of the classical computation theory such as quantum 1-way and 2-ways automata, quantum Tjuring machines etc. have been defined. It has been proved that some of these constructions are more powerful that their classical counterparts (see [2], page 150 for more details). The aim of this paper is to show how it is possible to "move" one of the classical computation theory constructions, namely quantum finite one-counter automata, to the "world of quantum computation", and compare it with deterministic and probabilistic one-counter finite automata.

The definition of introduced automata and its well-formedness conditions are quite similar to that of 2-way quantum automata in [1] and [2]. We introduce also a special case of QF1CA called simple, with easier conditions, that allow construct such automata easier than in general case. Some features of language recognition by the automata are considered.

Recognition of some languages with the introduced automata, namely $0^n10^n$, $0^n10^n10^n$, $0^n1^n$ and $0^l10^m10^n${exactly 2 of l,m,n are equal}, $0^l10^m10^n${(l=n or m=n) and $\neg$(l=m))} is shown. While first two are simply reversible counterparts for corresponding deterministic and probabilistic automata, the other use some quantum computation properties.

## 2. Classical one-counter automata

**Definition 1:** A one-counter deterministic finite automaton (1CDFA) *A* is specified by the finite (input) alphabet $\Sigma$, the finite set of states Q, the initial state $q_0$, the sets $Q_a \subset Q$ and $Q_r \subset Q$ of accepting and rejecting states, respectively, with $Q_a \cap Q_r = 0$, and the transition function $: Q \times \Sigma \times S \to Q \times \{\leftarrow \downarrow \rightarrow\}$, where S={0,1}.

There is a counter that can contain an arbitrary large integer value, it is 0 at the beginning of computation. $\leftarrow, \downarrow, \rightarrow$ respectively, decreases by one, retains the same and increases by one the value of the counter. S is defined to be 0 if and only if the value of the counter is equal to 0, otherwise S=1.

---

[*] Research supported by Grant No. 96.0282 from the Latvian Council of Science and Grant No. 1067 from the Sponsorship Programme of Latvian Fund of Education for Education, Science, and Culture.

The computation of 1CDFA on the input word x∈Σ* can be described as follows. We assume that word is written on the tape. The automaton (in the state q) reads a letter of the word written on the tape (σ), checks the value of the counter (s), finds an appropriate value of the transition function $(q, \sigma, s) \to q', d$. Then the state of the automaton changes to a new state q' and the value of the counter changes according to value of d as described above. Than the computation continues with the next letter of the word. After reaching of the end of the word if the automaton is in the accepting state, then the automaton accepts the word, if it is in rejecting state then the automaton rejects the word.

**Example 1:** 1CDFA can recognize, for example, the following language: $L_1$ $0^n 10^n$.

**Definition 2:** A one-counter probabilistic finite automaton (1CPFA) *A* is specified by the finite (input) alphabet Σ, the finite set of states Q, the initial state $q_0$, the sets $Q_a \subset Q$ and $Q_r \subset Q$ of accepting and rejecting states, respectively, with $Q_a \cap Q_r = 0$, and the transition function $\delta: Q \times \Sigma \times S \times Q \times \{\leftarrow \downarrow \rightarrow\} \to R^+$, where S={0,1} and δ satisfies the following condition:
$$\sum_{q',d} \delta(q, \sigma, s, q', d) = 1$$ for each q, q'∈Q, σ∈Γ, s∈{0,1}, d∈{←,↓,→}.

**Example 2:** It has been proven that C1PFA can recognize the language $L_2$ $0^n 10^n 10^n$ with probability 1-1/n, for each n∈N>=2. The basic idea of the automaton is that the probabilistic decision is made during the first step and one of the following n paths is chosen with equal probability. Each path is a deterministic automaton that accepts the word if it is in the form $0^i 10^j 10^k$ and an equation in the form a*i +b*j= (a+b) *k, where a, b ∈ N is satisfied. We can choose such a, b for each path that the equation can be satisfied at most in one path for any word, which is in form $0^i 10^j 10^k$ and does not belong to $L_2$. Thus if the word belongs to $L_2$ than the automaton accepts it with probability 1. If the word is not like $0^i 10^j 10^k$ than it is rejected with probability 1. If the word does not belong to $L_2$ but is like $0^i 10^j 10^k$ then it is rejected with probability at least 1-1/n

Now we can consider the quantum counterpart.

## 3. Quantum finite one-counter automata

**Definition 3:** A quantum finite one-counter automaton (MM-QF1CA) *A* is specified by the finite (input) alphabet Σ, the finite set of states Q, the initial state $q_0$, the sets $Q_a \subset Q$ and $Q_r \subset Q$ of accepting and rejecting states, respectively, with $Q_a \cap Q_r = 0$, and the transition function $\delta: Q \times \Gamma \times S \times Q \times \{\leftarrow \downarrow \rightarrow\} \to C_{[0,1]}$, where $\Gamma = \Sigma \cup \{\#, \$\}$ is the tape alphabet of A and symbols #, $ are endmarkers not in Σ, S={0,1}, and δ satisfies the following conditions (of well-formedness) for each $q_1, q_2, q' \in Q$, σ∈Γ, s∈{0,1}, d∈{←,↓,→}:
(1) Local probability and orthogonality condition:
$$\sum_{q',d} \delta^*(q_1, \sigma, s, q', d) \delta(q_2, \sigma, s, q', d) = \begin{cases} 1, & if \quad q_1 = q_2 \\ 0, & if \quad q_1 \neq q_2 \end{cases}$$

(2) Separability condition I

$$\sum_{q'} \delta^*(q_1,\sigma,s_1,q',\rightarrow)\delta(q_2,\sigma,s_2,q',\downarrow) + \sum_{q'} \delta^*(q_1,\sigma,s_1,q',\downarrow)\delta(q_2,\sigma,s_2,q',\leftarrow) = 0$$

(3) Separability condition II

$$\sum_{q'} \delta^*(q_1,\sigma,s_1,q',\rightarrow)\delta(q_2,\sigma,s_2,q',\leftarrow) = 0, \text{ where * denotes complex conjunctive}$$

Formally $A=(\Sigma, Q, q_0, Q_a, Q_r, \delta)$

In order to process an input word $x \in \Sigma^*$, we assume that the word is written on the tape in the form $w_x = \#x\$$. The definition of a counter and actions with it remain the same as for deterministic automaton.

We begin by proving that the evolution of a QF1CA $A$, satisfying these conditions, is unitary.

For an integer n let $C_n$ be the set of all possible configurations of $A$ for inputs of length n. The definition determines that at the n-th step automata reads n-th symbol of $w_x$, and before the n-th step the counter can contain value from -(n-1) up to n-1, So the configuration of A for each specific input x at each step can be uniquely determined by a pair (q,k) $q \in Q$ and $k \in [0, n-1]$, where q is the state of the automata and k is value of the counter.

A computation of A on an input x of length n corresponds to a unitary evolution in the underlying Hilbert space $H_{A,n}=l_2(C_n)$ (see [2] for more details). For each $c \in C_n$, $|c\rangle$ denotes the basis vector in $l_2(C_n)$, we'll use also $|q,k\rangle$. Each state in $H_{A,n}$ will therefore have a form $\sum_{c \in C_n} \alpha_c |c\rangle$ where $\sum_{c \in C_n} |\alpha_c|^2 = 1$.

The automaton A induces for any input $x \in \Sigma^n$ a linear operator $U_x^\delta$ that is defined for a configuration $(q,k) \in C_n$ $U_x^\delta |q,k\rangle = \sum_{q',d} \delta(q, w_{xi}, sign(k), q', d) |q', k+\mu(d)\rangle$, where $w_{xi}$ denotes i-th symbol of $w_x = \#x\$$, sign(k)= 0 if k=0 and 1 otherwise, $\mu(d)$ = -1(0)[1] if d=$\leftarrow(\downarrow)[\rightarrow]$. By linearity $U_x^\delta$ is extended to map any superposition of basis states.

**Theorem 1:** For any input string x the mapping $U_x^\delta$ is unitary if and only if the conditions (1) to (3) of **Definition 3** are satisfied.

Proof. To prove the Theorem, it is necessary to show that these conditions correspond to unitarity conditions of $U_x^\delta$ clearly $U_x^{\delta^*} \times U_x^\delta = E$, that can be rewritten as:

1. $\left\| U_x^\delta |q,k\rangle \right\| = \left\langle U_x^\delta |q,k\rangle \middle| U_x^\delta |q,k\rangle \right\rangle = 1$ for all configurations (q,k)

2. $U_x^\delta |q_1,k_1\rangle \perp U_x^\delta |q_2,k_2\rangle$ for all different configurations $(q_1,k_1)$ and $(q_2,k_2)$ (or the same
   $\left\langle U_x^\delta |q_1,k_1\rangle \middle| U_x^\delta |q_2,k_2\rangle \right\rangle = 0$ **(3.1)** and $\left\langle U_x^\delta |q_2,k_2\rangle \middle| U_x^\delta |q_1,k_1\rangle \right\rangle = 0$ **(3.2)** )

   where q, $q_1$, $q_2 \in Q$ and k, $k_1$, $k_2 \in [0,|x|+1]$.

$$\langle U_x^\delta |q_1,k_1\rangle | U_x^\delta |q_2,k_2\rangle\rangle = \sum_{q',d_1,d_2} \delta^*(q_1,w_{xi},\text{sign}(k_1),q',d_1)\delta(q_2,w_{xi},\text{sign}(k_2),q',d_2), \text{ where}$$

$k_1 + d_1 = k_2 + d_2$. Each member of the sum corresponds to the product of the amplitudes of $U_x^\delta |q_1,k_1\rangle^*$ and $U_x^\delta |q_2,k_2\rangle$ mapping to the same configuration $|q',k_1+d_1\rangle = |q',k_2+d_2\rangle$.

If conditions (1) to (3) of Definition 3 are satisfied, then

1. $\left\| U_x^\delta |q,k\rangle \right\| = \sum_{q',d} \delta(q,w_{xi},\text{sign}(k),q',d)^* \delta(q,w_{xi},\text{sign}(k),q',d) = 1$ when (1) is true for $q_1=q_2$

2. To show (3.1) and (3.2) we observe separately the following cases ($k_1 \le k_2$):

2.1. $k_1 = k_2$ ($q_1 \ne q_2$) → $(3.1) = \sum_{q',d} ä^*(q_1,w_{xi},\text{sign}(k),q',d)ä_2,w_{xi},\text{sign}(k),q',d) = 0$, when (1) is true for $q_1 \ne q_2$. When writing down (2.2) in this case we get the same sum as above.

2.2. $k_2 - k_1 > 2$ → $(3.1) = (3.2) = 0$. There is no such $|q',k'\rangle$ for which there is non zero amplitude in both $U_x^\delta |q_1,k_1\rangle$ and $U_x^\delta |q_2,k_2\rangle$ in this case, because the value of the counter can change at most by 1 at each step.

2.3. $k_2 - k_1 = 2$ → $(3.1) = \sum_{q'} \delta^*(q_1,w_{xi},\text{sign}(k_1),q',\rightarrow)\delta(q_2,w_{xi},\text{sign}(k_2),q',\leftarrow) = 0$, when (3) is true. $(3.2) = \sum_{q'} \delta(q_1,w_{xi},\text{sign}(k_1),q',\rightarrow)\delta^*(q_2,w_{xi},\text{sign}(k_2),q',\leftarrow)$. This expression differs from expression for (2.1) only by the opposite sign of imaginary part. Each member of the sums (3.1) and (3.2) differs as $(\_1 - \_1 i)(\_2 + \_2 i) = \_1\_2 + \_1\_2 + (\_1\_2 - \_1\_2)i$ and $(\_1 + \_1 i)(\_2 - \_2 i) = \_1\_2 + \_1\_2 - (\_1\_2 - \_1\_2)i$. So (3.2) in this case is 0 too.

2.4. $k_2 - k_1 = 1$ → $(3.1) = \sum_{q'} \delta^*(q_1,w_{xi},\text{sign}(k_1),q',\downarrow)\delta(q_2,w_{xi},\text{sign}(k_2),q',\leftarrow) +$

$+ \sum_{q'} \delta^*(q_1,w_{xi},\text{sign}(k_1),q',\rightarrow)\delta(q_2,w_{xi},\text{sign}(k_2),q',\downarrow) = 0$, when condition (2) is true.

Satisfaction of (3.2) can be shown like satisfaction of (3.2) in the previous part 2.3.

We see that these conditions mathematically are quite similar to those for 2-way quantum finite automata from [1] and [2] page 155, but logically they express quite different processes. For the automata with counter the position of the head is determined at each step and computation takes exactly the length of input word steps. But at the same time the changing of the counter value is not deterministic.

Conditions (1) - (3) are not very easy to test for the concrete automaton. So like the definition of simple 2-way QFA, see [1] and [2] page 156, we can define **simple QF1CA.**

**Definition 4:** A QF1CA is simple, if for each $\sigma \in \Gamma$, $s \in \{0,1\}$ there is a linear unitary operator $V_{\sigma,s}$ on the inner product space $l_2(Q)$ and a function $D: Q, \Gamma \rightarrow \{\leftarrow,\downarrow,\rightarrow\}$ such that for each $q \in Q$, $\sigma \in \Gamma$, $s \in \{0,1\}$

$$\delta(q,\sigma,s,q',d) = \begin{cases} \langle q'|V_{\sigma,s}|q\rangle & \text{if} \quad D(q',\sigma) = d \\ 0, & \text{else} \end{cases}$$

(where $\langle q'|V_{\sigma,s}|q\rangle$ denotes the coefficient of $|q'\rangle$ in $V_{\sigma,s}|q\rangle$).

**Theorem 2.** A simple QF1CA satisfies the well-formedness conditions (1) - (3) if and only if
$$\sum_{q'}\langle q'|V_{\sigma,s}|q_1\rangle^* \langle q'|V_{\sigma,s}|q_2\rangle = \begin{cases} 1 & \text{if} \quad q_1 = q_2 \\ 0 & \text{else} \end{cases}, \text{ for each } \sigma \in \Gamma, s \in \{0,1\}. \text{ That holds if and only if}$$
every operator is unitary.

Proof: we can simply rewrite well-formedness conditions:
$$\Sigma_{q',d}\delta^*(q_1,\sigma,s,q',d)\delta(q_2,\sigma,s,q',d) = \sum_{q'}\delta(q,\sigma,s,q',D(q',\sigma))^* \delta(q,\sigma,s,q',D(q',\sigma)) + 0 =$$

1)
$$= \sum_{q'}\langle q'|V_{\sigma,s}|q\rangle^* \langle q'|V_{\sigma,s}|q\rangle = \begin{cases} 1, & \text{if} \quad q_1 = q_2 \\ 0, & \text{if} \quad q_1 \neq q_2 \end{cases}$$

2) – 3) Both separability conditions are satisfied because $D(q',\sigma)$ can be equal only to one of the $\leftarrow, \downarrow, \rightarrow$ for all $\delta$ in the sum. But each member of the sums (2) (3) has two multipliers with different d ($\leftarrow, \downarrow$ and $\downarrow, \rightarrow$ in (2) and $\leftarrow, \rightarrow$ in (3)). So each member of the sum has at least one of the multipliers equal to 0. Thus the whole sum is 0 too.

We can use these considerations also to prove the theorem in the opposite direction.
Now we can define how computation of QF1CA proceeds.

## 4. Language recognition for QF1CA

Acceptance and rejection can be defined for QF1CA in some ways similarly as for push-down quantum finite automata (PDQFA) (see [4]).

An observable used in QF1CA is defined like the description of observation of 2-way quantum automata [2], page 158. The difference is that here acceptance can be defined in 3 different ways 1) acceptance both by state and zero value of the counter 2) acceptance by zero value of the counter 3) acceptance by state:

For each input word x with n=|x| and a QFA $A=(\Sigma, Q, q_0, Q_a, Q_r, \delta)$ let $C_n^a$ = {(q,k) | (q,k)$\in C_n$, 1) q$\in Q_a$, k=0 2) k=0 3) q$\in Q_a$,}, $C_n^r$ = {(q,k) | (q,k)$\in C_n$, 1) q$\in Q_r$ 2) q$\in Q_r$ k<>0 3) q$\in Q_r$ }, and $C_n^- = C_n - C_n^a - C_n^r$. Let $E_a$, $E_r$ and $E_-$ be the subspaces of $l_2(C_n)$ spanned by $C_n^a$, $C_n^r$ and $C_n^-$ respectively.

The "computational observable" $O$ corresponds to the orthogonal decomposition $l_2(C_n)$ = $E_a \oplus E_r \oplus E_-$. The outcome of any observation will be either "accept" ($E_a$) or "reject" ($E_r$) or "non-terminating" ($E_-$).

The language recognition by A is now defined as follows: for an $x \in \Sigma^*$ as the input is used $w_x$=#x$, and computation starts in the state $|q,0\rangle$ - counter is set to 0. For each letter from $w_x$ operator $U_x^\delta$ is applied to current state and the resulting state is observed using the computational observable $O$ defined above. After it the state collapses into $E_a$, or $E_r$ or $E_-$. If "non-terminating" state is observed than computation continues with next letter. The probability of

the acceptance, rejection and non- terminating at each step is equal to the square of amplitude of new state for the corresponding subspace. Computation stops either after halting state is observed or word is proceeded.

Another approach is to define that measurement will be made only after applying operators $U_x^\delta$ for each letter from $w_x$ to initial state.

For these two different approaches of definition of 1-way QFA it has been proved (see [2] page 152) that MO (measure one) QFA can be simulated by MM (measure many) QFA, but opposite is not true. For the QF1CA it can be shown too.

**Theorem 3**. MO QF1CA $A=(\Sigma, Q, q_0, Q_a, Q_r, \delta)$ automaton accepting by state or by state and zero value of counter can be simulated by the MM QF1CA $A_1=(\Sigma, Q_1, q_0, Q_{a1}, Q_{r1}, \delta_1)$ with the same acceptance type.

Proof. To show that MO QF1CA A can be simulated by MM QF1CA $A_1$, it is sufficient to show that no measurement in $A_1$ that is made before measurement after \$ disturbs the actual state of the automaton. No disturb means that amplitudes for $E_a$, and $E_r$ are both 0.

The states of $A_1$ are constructed by adding the same count of states to Q. $Q_1=Q\cup Q'$. Accepting and rejecting states for the new automaton are defined to be only from Q'. $Q_a'=\{q'_i|q_i\in Q_a\}$, $Q_r'=\{q'_i| ,q_i\in Q_r\}$.

Transition function is $\delta_1$ is defined as $\delta_1(q_1,\sigma,s,q_2,d) = \delta(q_1,\sigma,s,q_2,d)$ for all $q_1,q_2\in Q$, $\sigma \in \Gamma|\$$, $s\in\{0,1\}$, $d\in\{\leftarrow,\downarrow,\rightarrow\}$ and $\delta_1(q_1,\$,s,q_2',d) = \delta(q_1,\$,s,q_2,d)$. Remaining part of $\delta$ must be defined so that $\delta_1$ satisfies well-formedness conditions (1) (3).

It can be done by defining $\delta_1$ all $q'_1\in Q'$, $\sigma \in \Gamma|\$$, $s\in\{0,1\}$, $d\in\{\leftarrow,\downarrow,\rightarrow\}$ $\delta_1(q_1',\sigma,s,q_1',\downarrow) = 1$, $\delta_1(q_1',\$,s,q_2,d) = \delta(q_1,\$,s,q_2,d)$. $\delta_1(q_1',\$,s,q_2',d) = 0$ and $\delta(q_1,\sigma,s,q_2,d) = 0$.

Configuration of $A_1$ is equal to that of A during processing of the same arbitrary input word before the final \$, due to definition of $\delta_1$, equal to $\delta$. No measurement disturbs the state of $A_1$ during it, because all $q\in Q$ are non-halting. When processing \$ the final measurement of $A_1$ accepts and rejects the word with the same probability as A, due to $\delta_1(q_1,\$,s,q_2',d) = \delta(q_1,\$,s,q_2,d)$ and definition of accepting and rejecting states for $A_1$.

But for the acceptance by zero counter only this construction does not work, because of halting when counter is equal to 0. The opposite statement that MM QF1CA can be simulated by MO QF1CA is false. So we cosider only MM QF1CA further.

The acceptance of language L for QF1CA can be defined in the same way as for other classes of automata (see [2], page 152 for 1-Way QFA). A QF1CA is said to accept a language L with probability $\frac{1}{2}+\varepsilon, \varepsilon>0$, if it accepts (halts in the accepting state after measurements) with probability at least $\frac{1}{2}+\varepsilon$ and rejects any $x\notin L$ with probability at least $\frac{1}{2}+\varepsilon$.

## 5. Negative values of the counter.

It is not clear which definition of the counter is more natural whether counter that holds only non-negative values (5.1) or counter that holds arbitrary integer values (5.2).

For C1PFA and C1DFA it has been proved that any automaton of kind (5.2) can be simulated by another automaton satisfying definition (5.1). A similar result can be proved for QF1CA.

**Theorem 4:** Each QF1CA $A=(\Sigma, Q, q_0, Q_a, Q_r, \delta)$ that allows transition of kind $\delta(q_1,\sigma,0,q_2,\leftarrow) \neq 0$ can be simulated with another QF1CA $A_1=(\Sigma, Q_1, q_0, Q_{a1}, Q_{r1}, \delta_1)$ without such transitions.

Proof: $Q_1$ will be defined as adding to $Q$ a new duplicate of states $Q'$. $Q_{a1}=\{q_i$ and $q'_i \mid q_i \in Q_a\}$, $Q_r'=\{q_i$ and $q'_i \mid q_i \in Q_r\}$.

Transition function $\delta_1$ is defined as:

$\delta_1(q_1,\sigma,s,q_2,d) = \delta(q_1,\sigma,s,q_2,d)$ if $s \neq 0$ or $d \neq \leftarrow$; $\delta_1(q'_1,\sigma,s,q'_2,d) = \delta(q_1,\sigma,s,q_2,-d)$ if $s \neq 0$ or $d \neq \rightarrow$; $\delta_1(q_1,\sigma,0,q'_2,\rightarrow) = \delta(q_1,\sigma,0,q_2,\leftarrow)$ $\delta_1(q'_1,\sigma,0,q_2,\rightarrow) = \delta(q_1,\sigma,0,q_2,\leftarrow)$ for all $q_1,q_2 \in Q$, $q'_1,q'_2 \in Q'$, $\sigma \in \Gamma$, $s \in \{0,1\}, d \in \{\leftarrow,\downarrow,\rightarrow\}$. All the other values of $\delta_1$ are defined as 0.

We can rewrite conditions (1) (3) and see that they are satisfied for the new automaton. It is easy to see that computing of $A_1$ differs from one of $A$ by $A_1$ contains configurations $|q',k\rangle$ instead of configurations $|q',-k\rangle$ A for each $k<0$.

## 6. Languages recognized by QF1CA

**Example 3.** A simple QF1CA recognizing $L_1$ $0^n 1 0^n$.

$Q=\{q_0,q_1,q_2,q_a,q_r,q_{r2}\}$ $Q_a=\{q_a\}$, $Q_r=\{q_r\}$. $q_0$ is the initial state. The transition function is specified by transitions:

$V_{\#,0}|q_0\rangle=|q_1\rangle$  $V_{0,s}|q_1\rangle=|q_1\rangle$  $V_{1,s}|q_1\rangle=|q_2\rangle$  $V_{0,s}|q_2\rangle=|q_2\rangle$  $V_{\$,s}|q_1\rangle=|q_r\rangle$  $V_{\$,1}|q_2\rangle=|q_r\rangle$,

$V_{\$,0}|q_2\rangle=|q_a\rangle$ (where $V_{\sigma,s}|q\rangle$ defines transition function for $\sigma \in \Gamma$, $s \in \{0,1\}$, $q \in Q$) and the remaining transitions are defined arbitrary so that unitarity requirements are satisfied.
$D(q_1,\#)=\downarrow$, $D(q_1,0)=\rightarrow$, $D(q_2,0)=\leftarrow$, $D(q_2,1)=\downarrow$, $D(q_a,\sigma)=\downarrow$, $D(q_r,\sigma)=\downarrow$, $D(q_{r2},\sigma)=\downarrow$

So the automaton starts from $q_0$ and moves to $q_1$. While reading 0 it remains in $q_1$ and increases counter, after 1 it moves to $q_2$ and then while reading 0 remains in $q_2$ and decreases counter. If the counter is 0 when state is $q_2$ and 0 is read, then the word is rejected, when $ is read denoting the end of the word if counter is 0 then the word is accepted else rejected.

Actually it is a deterministic reversible automaton. The automaton accepts the word in $L_1$ with probability 1 and rejects words that are not in $L_1$ with probability 1.

This automaton is in accepting state only when counter value is 0 before reading $. So any of the acceptance types can be used.

This automaton can be easily modified to work only with nonnegative counter value (transition $V_{0,s}|q_2\rangle=|q_2\rangle$ must be substituted by $V_{0,1}|q_2\rangle=|q_2\rangle$, $V_{0,0}|q_2\rangle=|q_r\rangle$, non specified transitions can be still defined arbitrary so that unitarity requirements are satisfied).

**Example 4.** A simple QF1CA recognizing $0^n 1^n$ by state and zero value.

The construction is slightly modified construction for 1-way QFA recognizing $0^i1^j$ given in [3] - further 1QFA($0^i1^j$).

$Q=\{q_0,q_1,q_2,q_a,q_r,q_{r2}\}$ $Q_a=\{q_a\}$, $Q_r=\{q_r\}$. $q_0$ is the initial state. The transition function is specified by transitions

$V_{\#,0}|q\rangle = \sqrt{1-p}\,|q_1\rangle + \sqrt{p}\,|q_2\rangle$

$V_{0,s}|q_1\rangle = (1-p)|q_1\rangle + \sqrt{p(1-p)}\,|q_2\rangle + \sqrt{p}\,|q_r\rangle$  $V_{0,s}|q_2\rangle = \sqrt{p(1-p)}\,|q_1\rangle + p|q_2\rangle + \sqrt{1-p}\,|q_r\rangle$

$V_{1,s}|q_1\rangle = |q_r\rangle$      $V_{1,1}|q_2\rangle = |q_2\rangle$      $V_{1,s}|q_1\rangle = |q_r\rangle$      $V_{1,0}|q_2\rangle = |q_{r2}\rangle$

$V_{\$,s}|q_1\rangle = |q_r\rangle$      $V_{\$,1}|q_2\rangle = |q_{r2}\rangle$      $V_{\$,s}|q_1\rangle = |q_r\rangle$      $V_{\$,0}|q_2\rangle = |q_a\rangle$

and the remaining transitions are defined arbitrary so that unitarity requirements are satisfied.

    $D(q_1, \#) = \downarrow$ $D(q_2, \#) = \downarrow$      $D(q_1, 1) = \leftarrow$, $D(q_2, 1) = \leftarrow$

    $D(q_1, 0) = \rightarrow$, $D(q_2, 0) = \rightarrow$      $D(q_a, \sigma) = \downarrow$ $D(q_r, \sigma) = \downarrow$

How the computation of the automaton proceeds is shown in [3], there are added only counter changes, that ensure recognition of equality of the count 0 and 1.

The automaton can be in accepting state only when counter is 0. Thus any of the acceptance types can be used. This automaton accepts $0^n1^n$ with probability not less than 1QFA($0^i1^j$) accepts $0^i1^j$ and rejects all other words with probability not less than 1QFA($0^i1^j$) rejects word not in $0^i1^j$. In [2] it is shown that the total probability p of 1QFA($0^i1^j$) is the root of the equation $p=1-p^3 = 0.68\ldots$

**Example 5.** A simple QF1CA recognizing $L_2$ $0^n10^n10^n$ by state and zero value.

Simple QF1CA can be defined similar to 1CPFA of **Example 2.** $V_{\#,0}|q_0\rangle = \sum_{1..n}\frac{1}{\sqrt{n}}|q_i\rangle$

$D(q_i,\#) = \downarrow$. All the other transitions are strictly deterministic for each of $q_1$, $q_n$. For example further transitions that correspond to a=1 b=0 from Example 2 look like:

$V_{0,s}(q_1)=q_1$, $D(q_1,0)= \rightarrow$;      $V_{1s}(q_1)=q_2'$, $D(q_2',1)= \downarrow$ ;

$V_{0,s}(q_2')= q_2'$, $D(q_2',0)= \downarrow$;      $V_{1,s}(q_2')=q_3'$, $D(q_3',1)= \downarrow$;

$V_{0,s}(q_3')= q_3'$, $D(q_3',0)= \leftarrow$; $V_{\$,0}(q_3')=q_{acc}'$, $D(q_{acc}',1)= \downarrow$; $V_{\$,1}(q_3')=q_{rej}'$, $D(q_{rej}',1)= \downarrow$;

all the other transitions for each "path" can be specified to reject the word and keep $V_{\sigma s}$ unitary. Some additional rejecting states are necessary for it.

This automaton acts the same as corresponding 1CPFA. Thus the probability of recognizing $L_2$ by it is the same as for 1CPFA.

# 7. Recognizing $L_3$ and $L_4$.

**Theorem 5**. There is a QF1CA A that recognizes language $L_3 = 0^l10^m10^n$ ((l=n or m=n) and $\neg$(l=m)) by state and zero counter with probability $\frac{4}{7}$.

Proof. Let $V_{\#,0}|q_0\rangle = \sqrt{\frac{2}{7}}|q_1\rangle + \sqrt{\frac{2}{7}}|q_2\rangle + \sqrt{\frac{3}{7}}|q_a\rangle$ where $q_0$ is initial state, $q_1$, $q_2$ non-terminating states, $q_a$ accepting state. Transitions for $q_1$ and $q_2$ and $q_3$ for every $\sigma \in \Gamma\,|\,\$$ can be defined in such way, that each of them would be reversible but deterministic and the following conditions are satisfied:

1) If the word is not like $0^i10^j10^k$ than the word is rejected in each path before it reaches #

2) If the word is of that kind that no rejection or acceptance occur during the computation. The first path leads to the state $q_1'$ and counter equal to the m-n, the second – $q_2'$ and l-n (such a computation is easy to build if allow counter to be negative and use transitions like one from the Example 5).

So before reading \$ word is accepted with p=$\frac{3}{7}$ (during the first step), rejected if it is not of kind $0^i 10^j 10^k$ with probability $1-\frac{3}{7}=\frac{4}{7}$, and is in the superposition $|q'\rangle = \sqrt{\frac{2}{7}}|q'_1, l-n\rangle + \sqrt{\frac{2}{7}}|q'_2, m-n\rangle$ otherwise.

We can specify the transitions for \$ as follows.

$V_{\$,0}|q'_1\rangle = \frac{1}{\sqrt{2}}|q_a\rangle - \frac{1}{\sqrt{2}}|q_r\rangle$     $V_{\$,0}|q'_2\rangle = -\frac{1}{\sqrt{2}}|q_a\rangle - \frac{1}{\sqrt{2}}|q_r\rangle$, D(\$,$q_a$)=↓

$V_{\$,1}|q'_1\rangle = -\frac{1}{\sqrt{2}}|q_a\rangle + \frac{1}{\sqrt{2}}|q_r\rangle$     $V_{\$,1}|q'_2\rangle = \frac{1}{\sqrt{2}}|q_a\rangle - \frac{1}{\sqrt{2}}|q_r\rangle$

After proceeding \$ we get:

1. If l=m=n then the state after reading \$ is $(\frac{1}{\sqrt{7}} - \frac{1}{\sqrt{7}})|q'_a, 0\rangle + (-\frac{1}{\sqrt{7}} - \frac{1}{\sqrt{7}})|q'_r, 0\rangle$. The word is rejected with p=$\left(2\sqrt{\frac{1}{7}}\right)^2 = \frac{4}{7}$. which is at the same time the total probability of rejection.

2. If all l, m, n are different, then the word is rejected with probability p=$\frac{4}{7}$, because the counter value is not 0 for both $|q'_1, l-n\rangle$.

3. (l=n)and($\neg$(l=m)) then the state after reading \$ is $\frac{1}{\sqrt{7}}|q_a, 0\rangle - \frac{1}{\sqrt{7}}|q_r, 0\rangle - \frac{1}{\sqrt{7}}|q_a, m-n\rangle - \frac{1}{\sqrt{7}}|q_r, m-n\rangle$. So the word is accepted at the \$ with p= $\frac{1}{\sqrt{7}^2} = \frac{1}{7}$ and total probability of accepting of the word p=$\frac{1}{7} + \frac{3}{7} = \frac{4}{7}$

4. (m=n)and($\neg$(l=m)). The same as shownin the previous item.

So the automaton recognizes $L_3$ with probability $\frac{4}{7}$.

**Theorem 6** There is a QF1CA A that recognizes language $L_3$ = $0^l 10^m 10^n$ (l=m & m<>n)or (l=n & m<>n) or (n=m & l<>n) with probability 1/2+$\epsilon$.

Proof. Similar to one for the Theorem 5. The difference is that now 3 paths must be defined and the tranition function for \$ must be chosen such that annihilation of amplitudes of the accepting state can be achieved when l=n=m.

## 8. Open problems

Open problems the author works on can be divided into 2 groups:
1. Problems connected with different definitions of QF1CA.

2. Problems connected with the power of the automata.

The problem from the first group that can be of interest is:

**Problem 1.1**. Are all acceptance types equal in the sense that if language can be recognized with automata with one acceptance type, it will also be recognized with all the others?

Acceptance types described in the paper are 1) acceptance both by state and zero value of the counter 2) acceptance by zero value of the counter 3) acceptance by state only.

The author believes that there is no straightforward way for an automaton with one type of acceptance to be modified to get another automaton recognizing the same languages with another type of acceptance.

The main difference from the classical automata is that during each observation, halting part of the state vector (the part that corresponds to the accepting and rejecting subspaces in the measurement basis) disappears. Each of the described acceptance types has its own accepting subspace which is different from the others. The impact of the fact that no parts of the state disappear during measurement of the future transformations is not easily predictable.

Thus another similar problem can be defined:

**Problem 1.2**. Are other definitions of the automata' rejectance, which can change language recognition for the QF1CA, possible?

The second group problems are more fundamental.

**Problem 2.1**. Find description for the class of languages recognizable by QF1CA.

It has been not proved whether this kind of automata can recognize only the subset of the languages recognizable by the corresponding probabilistic automata or some other languages too. So a simpler arises:

**Problem 2.2**. Find the language that can be recognized by the QF1CA and can not be recognized by PF1CA.

On the one hand QF1CA is one way automata. And it is known that 1-way quantum automata can only recognize the subset of languages recognizable by 1-way deterministic automata.

On the other hand we have seen that computational model of QF1CA is very close to the one of 2-way quantum automata. And 2QFA can recognize more than probabilistic 2-way automata.

The main problem is that computation of any word takes exactly the length of the word steps and the value of the counter can change only by 1 at one step. So it seems impossible to "reset" the counter value, when it can be desired.

So it is difficult to achieve quantum interference, the situation where the advantages of quantum computing arise. Such situation was shown in Examples 4 and 5 when equality of both $m=n$ and $l=m$ lead to the elimination of the amplitude of the accepting state. At the same time when each $m,n,l$ were different, each of the base configurations of the state of the automaton corresponded to a different value of the counter and so no effect of quantum interference could occur.